\documentclass[final,5p,times,twocolumn]{elsarticle}

\usepackage{amssymb}
\usepackage{mathrsfs}
\usepackage{bm}
\newcommand{\V}[1]{\bm{#1} } 

\newcommand{\Ave}[1]{\left\langle {#1} \right\rangle} 

\newcommand{\mC}{\mathbb{C}}
\newcommand{\mR}{\mathbb{R}}
\newcommand{\mN}{\mathbb{N}}
\newcommand{\lb}{\left(}
\newcommand{\rb}{\right)}

\newcommand{\lsb}{\left[}
\newcommand{\rsb}{\right]}

\journal{Physica E}

\begin{document}

\begin{frontmatter}



\title{Zero-Temperature Complex Replica Zeros of the $\pm J$ Ising Spin Glass\\ on Mean-Field Systems and Beyond}


\author[label1]{Tomoyuki Obuchi}
\ead{obuchi@spin.ess.sci.osaka-u.ac.jp}
\author[label2]{Yoshiyuki Kabashima}
\author[label3]{Hidetoshi Nishimori}
\author[label3]{Masayuki Ohzeki}

\address[label1]{Department of Earth and Space Science, Faculty of Science, Osaka University, Toyonaka 560-0043, Japan}
\address[label2]{Department of
Computational Intelligence and Systems Science,
 Tokyo Institute of Technology, Yokohama 226-8502, Japan}
\address[label3]{Department of
Physics, Tokyo Institute
of Technology,  Tokyo 152-8551, Japan}

\begin{abstract}
Zeros of the moment of the partition function $[Z^n]_{\V{J}}$ with respect to complex $n$ are investigated in the zero temperature limit $\beta \to \infty$, $n\to 0$ keeping $y=\beta n \approx O(1)$. 
We numerically investigate the zeros of the $\pm J$ Ising spin glass models on several Cayley trees and hierarchical lattices and compare those results.  
In both lattices, the calculations are carried out with feasible computational costs by using recursion relations originated from the structures of those lattices. The results for Cayley trees show that a sequence of the zeros approaches the real axis of $y$ implying that a certain type of analyticity breaking actually occurs, although it is irrelevant for any known replica symmetry breaking.  
The result of hierarchical lattices also shows the presence of analyticity breaking, even in the two dimensional case in which there is no finite-temperature spin-glass transition, which implies the existence of the zero-temperature phase transition in the system. 
A notable tendency of hierarchical lattices is that the zeros spread in a wide region of the complex $y$ plane in comparison with the case of Cayley trees, 
which may reflect the difference between the mean-field and finite-dimensional systems. 
\end{abstract}


\end{frontmatter}


\section{Introduction}
\label{sec:intro}
Disordered systems are one of the challenging problems in 
statistical physics.
Especially, spin glasses 
have been investigated for a long time 
as an ideal and nontrivial problem treating disorder. 
One of the most important approaches in the spin-glass theory is 
the replica method. 
This method has provided both profound concepts
and useful calculation techniques in the theory, which 
promoted the expansion of the spin-glass theory to 
other disciplines after successful construction of a description 
of spin glasses \cite{SPIN,STAT,SLOW,INFO}. 

A characteristic property of disordered systems is its sample fluctuation of the thermodynamic quantities \cite{Bouchaud:97,Parisi:08,Nakajima:08}. 
In the framework of the replica method, the fluctuation, 
which is reflected in higher-order cumulants of the free energy, 
is essentially utilized to calculate the typical free energy. 
This is actually implemented by an assessment scheme of 
the cumulant generating function $\phi(n)$ defined as follows;
\begin{eqnarray}
\phi(n)=\lim_{N\to \infty}\frac{1}{N}\log [Z^n]_{\V{J}},
\end{eqnarray} 
where the brackets $[\cdots]_{\bm{J}}$ denote the average over the 
quenched disorder in the system. 
The typical free energy $f=-\lim_{N\to \infty }[\log Z]_{\V{J}}/(\beta N)$ is obtained from $\phi(n)$ 
as 
$-\beta f=\lim_{n\to 0}(\partial\phi(n))/(\partial n)$
, and any high-order cumulants  
can be similarly derived from higher-order derivatives of $\phi(n)$ with respect to $n$. 

In the usual replica framework, 
to obtain the full functional form of $\phi(n)$, 
we use the analytical continuation from $n\in \mN$ to $n\in \mR$ (or $\mC$).
This is because the exact assessment of $[Z^n]_{\V{J}}$ for $n \in \mR$ 
is generally infeasible except for a few solvable models.
However, this procedure of the replica method causes a problem: Some analyticity breaking can generally occur in $\phi(n)$ due to the limit $N \to \infty$, 
which is essentially incompatible with the analytic continuation used 
in the replica method.  
This means that the expression analytically continued from 
$n \in \mN$ to $n \in \mR$ 
will lead to an incorrect solution for the limit $n \to 0$ if the 
breaking of analyticity occurs in the region $0 < n < 1$.
To recover the correct solution of $\phi(n)$, 
in such cases, 
we need to know the properties of the analyticity breaking of $\phi(n)$ 
and to modify the solution according to the details.
This provides a motivation to develop a method directly investigating 
the analyticity breaking of $\phi(n)$ with respect to $n$.

This type of transitions of $\phi(n)$ is considered to be related 
to the replica symmetry breaking (RSB) in the Parisi scheme 
\cite{Parisi:80-1,Parisi:80-2},
which is considered to be exact for a wide variety of spin-glass models in 
the limit $n\to 0$ 
(it gives the exact result for the Sherrington-Kirkpatrick model
\cite{Sherrington:75,Talagrand:06}). 
Actually, 
for a variation of the discrete random energy model, 
it is shown that
the analyticity breaking of $\phi(n)$
actually occurs and is relevant to the one-step RSB (1RSB) 
\cite{Ogure:04,Ogure:09-1}.
This fact again motivates us 
to investigate the analytic behavior of $\phi(n)$ 
and to examine the 
phase transitions occurring in the region $n\geq 0$.

Under these motivations, we provide a method to investigate the 
analyticity breaking of $\phi(n)$, 
based on the Yang-Lee description of phase transitions \cite{Yang:52}.
In particular, 
 we observe the zeros of $[Z^n]_{\V{J}}$ with respect 
to complex replica number $n$, which will be referred to 
as ``replica zeros'' (RZs).

For the discrete random energy model mentioned above, 
this strategy successfully characterizes the 1RSB transition accompanied 
by a singularity of the large deviation rate function of the free energy 
$f$ \cite{Ogure:09-2}. 
On the other hand for the infinite-step of RSB (FRSB), 
a possibility that 
the RZs cannot characterize the FRSB is suggested \cite{Obuchi:09}, 
according to an argument based on the RZs of the $\pm J$ models on some tree-like systems and an analysis of the spin-glass susceptibility. 
These results require more detailed discussions about the relation between 
the analyticity of $\phi(n)$ and the RSB. 

Another interesting problem concerning the RZs is its application to 
finite-dimensional systems. 
It is still a subject of considerable discussion 
whether the RSB occurs or not in finite-dimensional systems. 
The RZs formulation can be a help to examine this problem.

For a concrete progress along the direction, we 
here treat hierarchical lattices \cite{Berker:79,Griffiths:82,Kaufman:84}.
Although it is known that 
the RSB is absent for spin-glasses 
on hierarchical lattices \cite{Gardner:84,Moore:98},
the dimensionality of these lattices can be tuned 
by changing a parameter and also
the renormalization group analysis 
gives the exact partition function.
These useful properties can make the hierarchical lattices be a productive first step to examine the finite-dimensional effects on the RZs of spin glasses.
 
This paper is organized as follows.
In the next section, 
we briefly summarize the formulation and the results for Cayley
trees in \cite{Obuchi:09}. 
In section 3, we provide a formulation and the RZs plots for the hierarchical lattices, and 
compare the results with those of Cayley trees. 
The last section is devoted to the conclusion.

\section{Formulation and results for Cayley trees}
\subsection{Basic formulation}
Our main objective is to solve the following equation with respect to $n$;
\begin{equation}
\lsb Z^n \rsb_{ \V{J} }=0.\label{eq:moment}
\end{equation}
This transcendental equation is, however, 
hard to solve for general temperatures. 
Then we here restrict ourselves to 
the zero temperature limit $\beta \to \infty$ 
involving $n\to 0$ and $\beta n\to y \approx O(1)$. 
In this limit, the relevant contribution to the partition function only 
comes from the ground state, and the RZs equation (\ref{eq:moment}) 
becomes 
\begin{equation}
\lsb e^{ -yE_{g} } \rsb_{ \V{J} }=0,\label{eq:moment2}
\end{equation}
where $E_{g}$ is the ground state energy.
Besides, we focus on the $\pm J$ models 
whose Hamiltonian and distribution of interactions without ferromagnetic bias are given by
\begin{eqnarray}
\mathcal{H}=-\sum_{ \Ave{i,j} }J_{ij}S_{i}S_{j}, \\
P( J_{ij} )=\frac{1}{2}\lb \delta(J_{ij}+1)+\delta(J_{ij}-1) \rb.
\end{eqnarray}
This limitation restricts the energy of the system to an integer value, which 
means that eq.\ (\ref{eq:moment2}) becomes a polynomial equation 
of $x=e^{y}$, and the computational cost to calculate the RZs equation
becomes significantly reduced.
\subsection{Results for Cayley trees}
For Cayley trees, we can efficiently calculate the moment $[Z^n]_{\V{J}}$
by combining the Bethe-Peierls approach and 
the replica method \cite{Obuchi:09}.  
Using this scheme, the RZs equation can be constructed and solved 
in a polynomial time with respect to the number of spins $N$ under appropriate boundary conditions. 
However, for Cayley trees, the number of spins and the degree of the
polynomial of $[Z^n ]_{\V{J}}$ exponentially increases as the characteristic 
length of the tree $L$ (distance between the central and boundary spins) 
grows. As a result, it is infeasible to solve the RZs equation
$[Z^n]_{\V{J}}=0$ for large $L$. 
The resultant plots for the $k=2$-body interacting Cayley tree with 
the coordination number $c=3$ 
and the $3$-body interacting Cayley tree with $c=3$  
are shown in fig.\ {\ref{fig:CTRZs}.
\begin{figure}[htbp]
\begin{tabular}{cc}
\hspace{-5mm}
\begin{minipage}[t]{0.5\hsize}
\begin{center}
 \includegraphics[height=38mm,width=45mm]{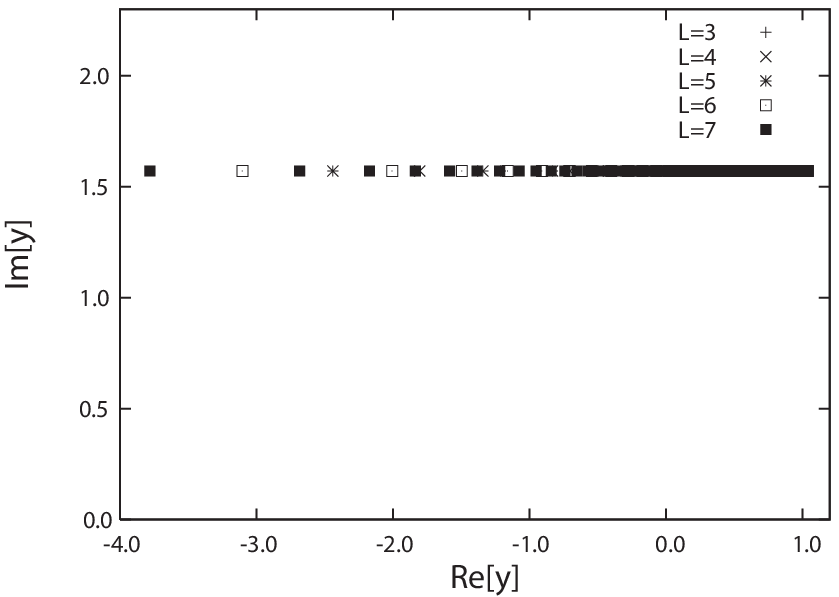}
\end{center}
\end{minipage}
\hspace{1mm}
 \begin{minipage}[t]{0.5\hsize}
\begin{center}
\includegraphics[height=38mm,width=45mm]{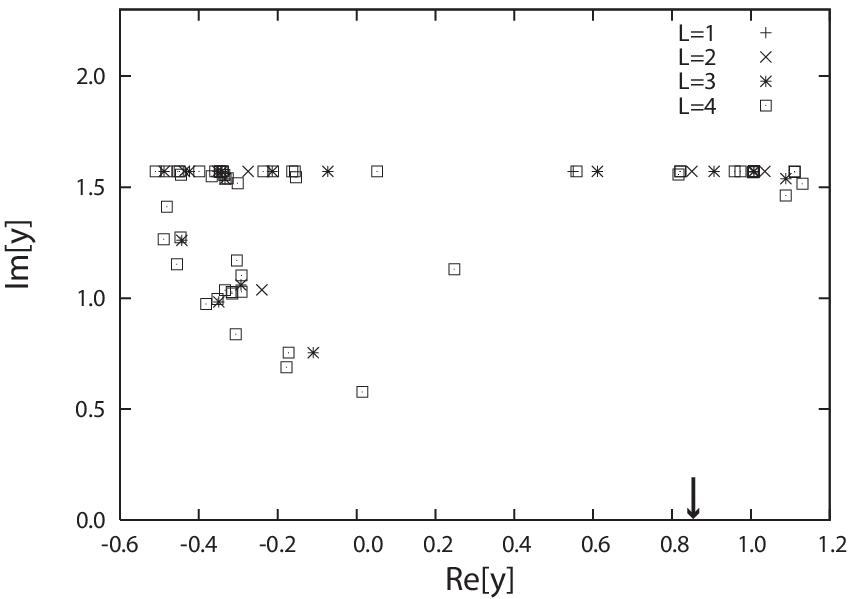}
\end{center}
\end{minipage}
\end{tabular}
\caption{RZs plots for $(k,c)=(2,3)$ and $(3,3)$ Cayley trees. 
All the zeros lie on a line ${\rm Im}(y)=\pi/2$ for the 
$(2,3)$ case (left panel) but
a sequence of zeros approach 
 the real axis as $L$ increases for the $(3,3)$ case (right panel). 
The arrow
 indicates the collision point expected from the analysis of 
the RS order parameter in the $L \to \infty$ limit.}
\label{fig:CTRZs}
\end{figure}
We can find two characteristic behavior in these plots. 
One is for the $(k,c)=(2,3)$ case (left panel) in which 
all the zeros lie on a line ${\rm Im}(y)=\pi/2$ and never 
reach the real axis.
That is in contrast to the other one for the $(3,3)$ case
(right panel) in which a sequence of zeros approaches the real 
axis as $L$ grows. 
These facts imply that analyticity breaking of $\phi(n)$ 
is absent for the $(2,3)$ case but present for $(3,3)$. 

An important point is whether 
the analyticity breaking in the $(3,3)$ case is 
related to the RSB or not. 
According to earlier studies, some RSB occurs in 
 the regular random graphs, 
which are known to share many properties
 with Cayley trees, 
of the same parameters 
$(k,c)=(2,3)$ and $(3,3)$ 
\cite{Bowman:82,Montanari:03}. 
This fact, combined with the apparent absence of the analyticity breaking 
in the left panel in fig.\ \ref{fig:CTRZs}, implies that 
the RZs of Cayley trees do not reflect any RSB. 
Hence, the analyticity breaking of $\phi(n)$ in the case $(3,3)$ should be 
considered to be a phase transition keeping the replica symmetry (RS). 
To see this, we observe the asymptotic behavior of the order parameter, 
which is given by the probability that the cavity field takes zero at a distance $L$ from the boundary $p_{L;0}$, and 
calculate $\lim_{L\to \infty}p_{L;0}=p^{*}$ \cite{Obuchi:09}. 
For the $(2,3)$ case, $p^{*}$ becomes a simple analytic function of $x=e^{y}$.
On the other hand,  for the $(3,3)$ case, $p^{*}$ shows 
non-analytic behavior and the result is given in fig.\ \ref{fig:apb3c3}.
\begin{figure}[htbp]
\begin{center}
   \includegraphics[height=38mm,width=55mm]{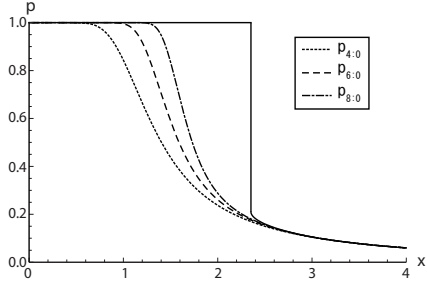}
 \caption{Asymptotic behavior of 
the order parameter of $(k,c)=(3,3)$ Cayley tree. 
The solid line denotes the $L \to \infty $ solution.
A finite jump of the order parameter $p^{*}=\lim_{L\to \infty}p_{L;0}$
occurs at $x\approx 2.35 \Leftrightarrow y \approx$ 0.85. 
}
 \label{fig:apb3c3}
\end{center}
\end{figure}
This figure shows a finite jump of the order parameter at $x\approx 2.35 \Leftrightarrow y\approx 0.85$. This singular point is indicated by an arrow on the right panel of fig.\ \ref{fig:CTRZs}. The approaching point of the RZs 
seems to agree with this singular point in observation by eye, 
which implies that the singularity indicated by the RZs corresponds to 
the singularity of the RS order parameter and does not reflect any RSB.



\section{Formulation and results for hierarchical lattices }
\subsection{Formulation for hierarchical lattices}
In this section, we treat 
the $\pm J$ models without ferromagnetic bias 
on hierarchical lattices.
A hierarchical lattice is consisted from unit cells \cite{Berker:79,Griffiths:82,Kaufman:84}. 
The structure of the cell determines the dimension of the system. 
Here we treat a simple cell consisting of two edge spins and $q$ 
inside spins (fig.\ \ref{fig:HL}).
Each pair of edge and inside spins is connected by the interaction.
\begin{figure}[htbp]
\begin{center}
   \includegraphics[height=25mm,width=80mm]{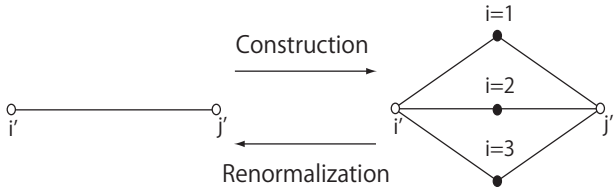}
 \caption{
A picture of a unit cell of $q=3$. White circles denote edge spins and 
black ones represent inside spins. 
}
 \label{fig:HL}
\end{center}
\end{figure}
The construction of a hierarchical lattice is performed by changing a bond 
to a unit cell, and the calculation of the partition function is conducted by 
the inverse procedure. 
The unit cell is renormalized to a bond between two edge spins by tracing out the inside spins. This yields the following equations 
expressed by the renormalization relations $R_{J}$ and $R_{F}$;  
\begin{eqnarray}
J'_{i' j'}=R_{J}(\{J_{ij} \}), \label{eq:rJ}\\
F_{i' j'}'=\sum_{i,j}F_{ij} +  R_{F}(\{J_{ij} \})\label{eq:rf},
\end{eqnarray}
where the set $\{ J_{ij} \}$ is for the bonds in the unit cell and 
$J_{i' j'}'$ denotes the renormalized bond between sites $i'$ and 
$j'$ of the renormalized system, and  
$F_{i' j'}'$ is the free energy of the system 
consisting of all the nested spins between $i'$ and $j'$.
Using the notations in fig.\ \ref{fig:HL}, the explicit forms for $R_{J}$ and $R_{F}$ are 
\begin{eqnarray}
R_{J}(\{J_{ij}\})=\sum_{i=1}^{q}\frac{1}{\beta}
\tanh^{-1}\lb
\tanh \beta J_{i i'}\tanh\beta J_{i j'}
\rb
=\sum_{i=1}^{q} \hat{J}_{i}
,\\
R_{F}(\{J_{ij}\})=-\frac{1}{\beta}\log\frac{1}{2^{q}} \frac{4 \cosh \lb \beta \sum_{i=1}^{q}\hat{J}_{i}\rb }{\prod_{i=1}^{q}4\cosh \beta \hat{J}_{i}}.
\end{eqnarray}
Using these relations, we can efficiently calculate the free energy of general Ising systems on the hierarchical lattices. 

In random systems on the hierarchical lattices, 
the probability distribution $P(J_{ij})$
 characterizes the behavior of the systems 
\cite{Mckay:82,Nobre:01}.
In the current problem, however, we need 
the joint probability distribution of the bond 
and free energy 
$P(F_{ij},J_{ij})$.  
The renormalized distribution $P'(F_{i' j'},J_{i' j'})$ 
is calculated from the original distribution $P(F_{i j},J_{ij})$
as 
\begin{eqnarray}
P'(F_{i' j'},J_{i' j'})=
\int \lb\prod_{i,j} dJ_{ij}dF_{ij}P(F_{ij},J_{ij}) \rb 
\nonumber \\
\times \delta \lb J'_{i' j'}-R_{J}(\{J_{ij} \}) \rb 
\delta \lb F'_{i' j'}-\sum_{i,j}F_{ij} - R_{F}(\{F_{ij} \}) \rb.
\label{eq:rP} 
\end{eqnarray}
Again we take the zero-temperature limit $\beta \to \infty$, $n\to 0$ keeping $\beta n \to y$.
In this limit, the free energy $F_{ij}$ becomes the ground state energy $E_{ij}$ and only takes an integer value, which is also the case for the bond $J_{ij}$.
This enables us to exactly perform the renormalization 
(\ref{eq:rP}) without numerical error.
Once we get the distribution $P(E_{ij},J_{ij})$, 
the RZs equation 
can be constructed by using the distribution of the energy $P(E_{ij})\equiv \int dJ_{ij}P(E_{ij},J_{ij})$ as 
\begin{equation}
\int dE_{ij} P(E_{ij})e^{-y E_{ij}}=0.\label{eq:RZeq}
\end{equation}
These equations (\ref{eq:rP}) and (\ref{eq:RZeq}), which 
enables us to exactly assess the RZs, 
constitutes the main result of this section.

\subsection{Results}
The procedures mentioned above, however, involve some difficulties. For hierarchical lattices, a characteristic length $L$ can be naturally defined as the number of hierarchy of the nested unit cells. As $L$ grows, the number of spins increases exponentially, which leads to the rapid growth of the support of $P(E_{ij})$ and 
the exponential increase of the degree of polynomial of the RZs equation. 
These cause difficulties in  
evaluating the convolution (\ref{eq:rP}) and 
solving the RZs equation (\ref{eq:RZeq}), which makes it hard 
to treat large size systems. 
These restrict the values of $q$ and $L$ to moderate values.  
For instance, in this paper, we investigate the cases $q=2$ and $3$ (the corresponding dimensions are $d=2$ and $2.58$, respectively) in the ranges $L\leq 5$ for $q=2$ 
and $L\leq 4$ for $q=3$. 
The resultant plots of RZs are given in fig. \ref{fig:HLRZs}.  
\begin{figure}[htbp]
\begin{tabular}{cc}
\hspace{-5mm}
\begin{minipage}[t]{0.5\hsize}
\begin{center}
 \includegraphics[height=38mm,width=45mm]{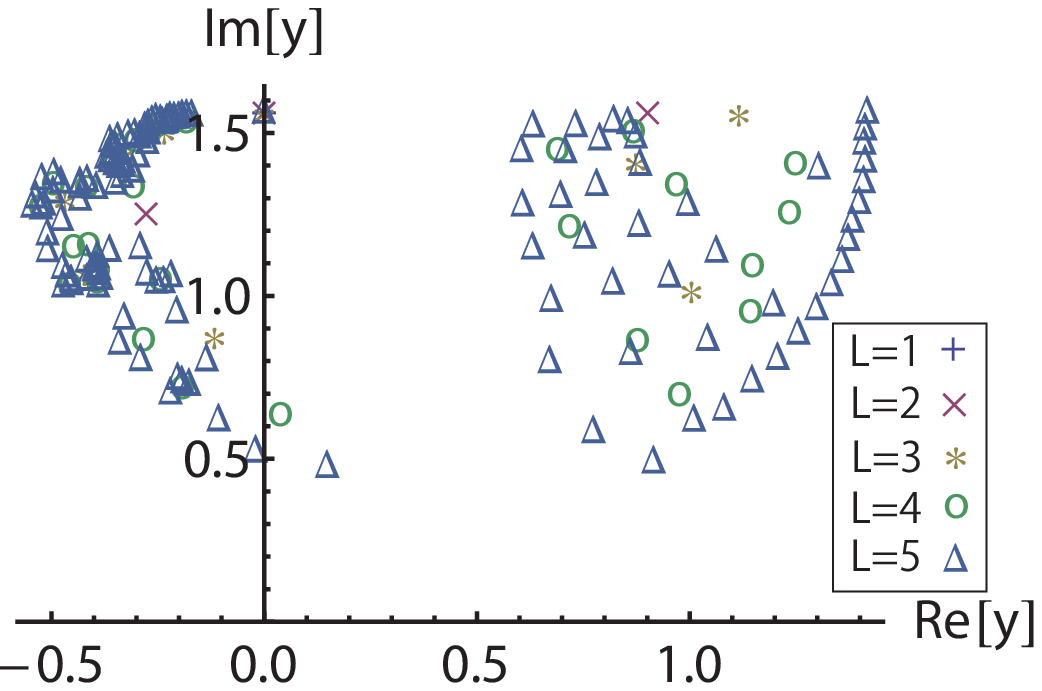}
\end{center}
\end{minipage}
\hspace{1mm}
 \begin{minipage}[t]{0.5\hsize}
\begin{center}
\includegraphics[height=38mm,width=45mm]{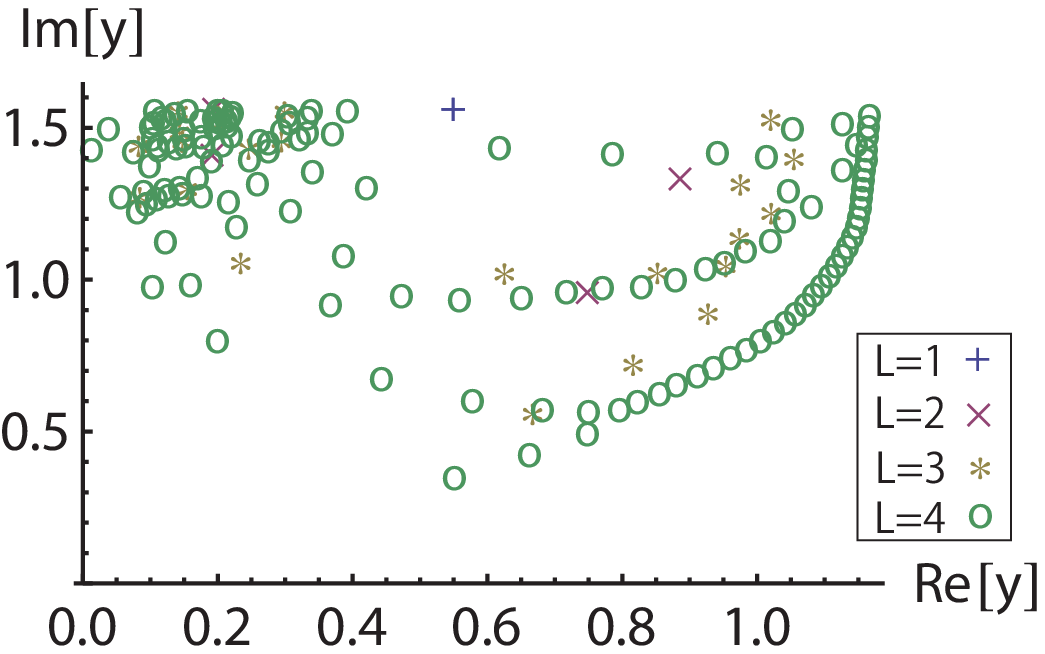}
\end{center}
\end{minipage}
\end{tabular}
\caption{The RZs of hierarchical lattices. The left panel is for $q=2$ and the right one is for $q=3$. In both panels, some zeros are spreading 
in a wide region, which is in contrast to the case of Cayley trees.
Also, sequences of zeros approach the real axis in both panels, which 
implies that a certain type of analyticity breaking occur in both cases.
}\label{fig:HLRZs}
\end{figure}

In this figure,
we can find that some sequences of the zeros approach 
the real axis of $y$ as $L$ grows in both $q=2$ and $3$ cases. 
For the two dimensional case $(q=2)$, 
those would be related to the zero-temperature spin-glass transitions, since the absence of the finite-temperature phase transition are clarified by several researches \cite{Nobre:01,Ohzeki:09}. 
To make this point clearer, we should treat the limit $L\to \infty$ as the case of Cayley trees, and now analyses along this direction are in progress.

Another interesting implication from fig.\ \ref{fig:HLRZs} is that 
the RZs are spreading in a wide region of the complex $y$ plane.
This is in contrast to the case of Cayley trees, which implies that the analyticity breaking of $\phi(n)$ in the hierarchical lattices might have different 
properties from those of Cayley trees. 
Since in general the continuous zeros distribution is related to the continuous singularities of the system \cite{Matsuda:10}, 
the RZs observed in fig.\ \ref{fig:HLRZs} 
may be related to extraordinary phase transitions,
although they would be different from the RSB transitions because 
the absence of the RSB for the present models is strongly suggested \cite{Gardner:84,Moore:98}. 
This point also needs more detailed analysis and 
a new approach to overcome the computational difficulty in assessing the RZs is desired.

\section{Conclusion}
In this paper, we have developed a formulation utilizing the zeros of the $n$th moment of the partition function, to directly examine the analyticity breaking of the cumulant generating function $\phi(n)$ 
appearing in the replica method, and  
applied it to the $\pm J$ models on Cayley trees and hierarchical lattices
in the zero temperature limit $\beta \to \infty$, $n\to 0$ keeping $\beta n\to y\approx O(1)$. 
The results imply the presence of analyticity breaking of the cumulant generating function $\phi(n)$ for both lattices, even in the two dimensional case of the hierarchical lattices in which there is no 
finite-temperature spin-glass transition.
Referring to other analytical properties of these systems \cite{Obuchi:09,Gardner:84,Moore:98}, we can reasonably conclude 
that those singularities revealed by the RZs are irrelevant to any known RSB.
Direct applications of the current scheme to systems exhibiting the RSB 
should be investigated in the future. 

In comparison with Cayley trees, the RZs of the hierarchical lattices tend to widely spread in the complex $y$ plane, which may be due to the finite-dimensional 
effect of the lattices. This point should also be clarified by further studies on the finite-dimensional spin glasses. 
At present, analytical approaches to finite-dimensional spin glasses are still rare. We hope that our current formulation becomes a useful analysis 
leading to further understanding of finite-dimensional spin glasses.

\section*{Acknowledgment}
This work was 
supported by the Japan Society for the Promotion of Science (JSPS) Research Fellowship for Young Scientists program, 
CREST(JST), the 21th Century COE Program `Nanometer-Scale Quantum Physics' 
and the Global COE Program `Nanoscience and Quantum Physics'
at Tokyo Institute of Technology, and by
the Grant-in-Aid for Scientific Research on
the Priority Area ``Deepening and Expansion of Statistical Mechanical
Informatics'' by  the Ministry of Education, Culture, Sports, Science
and Technology.













\end{document}